\documentclass[
  ,final            
  ]
  {aipproc}

\layoutstyle{6x9}
\usepackage{natbib}

\begin{document}

\title{Polarization characteristics of the Crab pulsar's giant radio pulses
at HFCs phases}

\classification{97.60.Gs}
\keywords      {pulsars, individual: Crab pulsar, giant radio pulses, polarization}

\author{A. S{\l}owikowska}{
  address={Nicolaus Copernicus Astronomical Center, Rabia\'nska 8, 87-100 Toru\'n, Poland}
}
\author{A. Jessner}{
  address={Max Planck Institute for Radioastronomy, Auf dem H\"ugel 69,
D-53121 Bonn, Germany}
}

\author{B. Klein}{
  address={Max Planck Institute for Radioastronomy, Auf dem H\"ugel 69,
D-53121 Bonn, Germany}
}

\author{G. Kanbach}{
  address={Max Planck Institute for Extraterrestrial Physics, Postfach 1312,
D-85741 Garching, Germany}
}

\begin{abstract}
We discuss our recent discovery of the giant radio emission
from the Crab pulsar at its high frequency components (HFCs)
phases and show the polarization characteristic of these pulses.
This leads us to a suggestion that there is no difference in the emission 
mechanism of the main pulse (MP), interpulse (IP) and HFCs. 
We briefly review the size distributions of the Crab 
giant radio pulses (GRPs) and discuss general characteristics of
the GRP phenomenon in the Crab and other pulsars.
\end{abstract}

\maketitle


\section{Introduction}
The occurrence of sporadic emission of intense pulses from the Crab pulsar
(PSR~B0531+21) has been known since its discovery in 1968 by
\citet*{Staelin1968}.  
\begin{figure}[h]
  \includegraphics[height=.36\textheight]{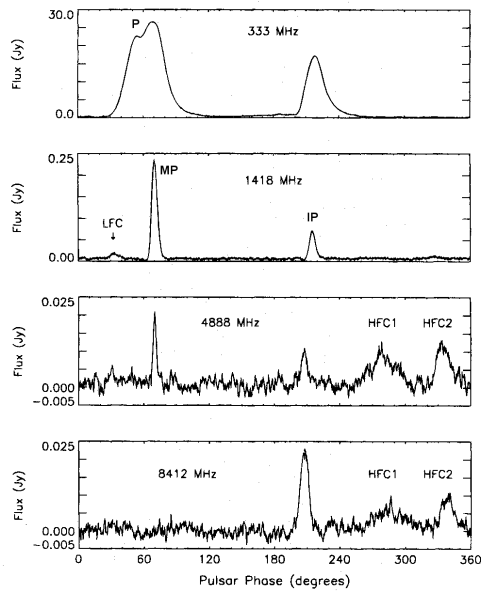}
  \caption{The aligned, average radio intensity profiles of the Crab
   pulsar obtained at 0.33, 1.4, 4.9 and 8.4~GHz with VLA
   showing several components: the main pulse (MP), interpulse (IP),
   precursor (P), low frequency component (LFC), and two broad
   high frequency components (HFC1, HFC2), waxing and waning with
   radio frequency.}
  \label{mh1996}
\end{figure}
It should be noticed that the pulsar would not have been discovered if it
had not showed the giant radio pulses phenomenon. For over 25 years,
only the Crab pulsar was known to emit giant radio pulses. Moreover,
for over 35 years, it was thought that only the main pulse and interpulse
exhibit the GRP phenomenon (e.g. \cite{Gower1972} - discovery of
GRPs at IP phase, \cite{Lundgren1995}, \cite{Sallmen1999}, \cite{Cordes2004}).
These two components have counterpart non-thermal emission from the infrared
to the gamma-ray energies. Until this year (\cite{Jessner2005}) the 
Crab's giant pulses were not seen in the radio precursor nor at the 
phases of the high radio frequency components (HFC1 and HFC2, see Fig.
\ref{mh1996}), that were first identified by \citet*{Moffett1996}. 
\begin{figure}
  \includegraphics[height=.50\textheight, angle=90]{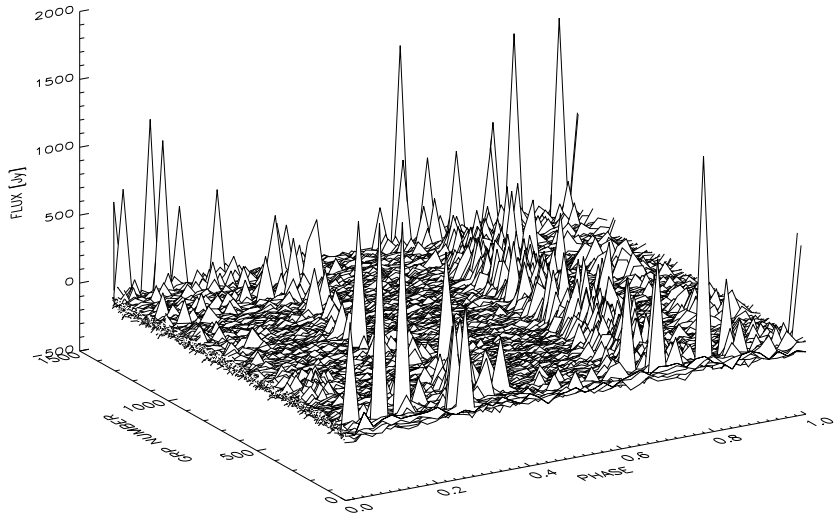}
  \caption{A train of strong giant pulses at 8.35 GHz, sum of the right and
  left handed circular polarisation signals. Results obtained during 
  6.7 hr of observation with the 100 m Effelsberg telescope.
Location of the Crab's pulsar radio components: HFC2: 0.05,
precursor: 0.2, MP: 0.3, IP: 0.68, HFC1: 0.9.}
  \label{grps}
\end{figure}

\section{Observations}
Observations with the Effelsberg 100-m radio telescope began on
25 November 2003 and ended on  28 November 2003. We used a  secondary
focus cooled  HEMT receiver with a center frequency of 8.35 GHz, providing 
two circularly polarized IF signals with a  system temperature of  25~K 
on both channels. With a sky temperature of 8~K, and a contribution of 
33~K from the Crab nebula, the effective system temperature was 70~K.
Two detection systems were used. First, a polarimeter with 1.1~GHz bandwidth 
detected total power of the left-hand and right-hand circularly polarized
signals (LHC and RHC respectively) as well as  $\cos \angle (\rm~LHC, RHC)$
and $\sin \angle (\rm~LHC, RHC)$. These four signals were then recorded by 
the standard  Effelsberg Pulsar Observation System (EPOS). With the Crab
pulsar's dispersion measure of $56.8~{\rm pc~cm^{-3}}$, we had an 
un--dedispersed time resolution of $t_{sample} = 890~\mu{\rm s}$,
while our sampling resolution was fixed at $\sim 640~\rm \mu s$.
EPOS was therefore set to continuously record data blocks containing
20 periods divided into 1020 phase bins for all four signals.
The minimum detectable flux per bin was $\Delta S_{min} = 0.117~{\rm Jy}$. 
Considering the dispersion pulse broadening, the detection limit for a
single GRP of duration $\tau_{grp} = 3~\mu{\rm s}$
would be $\Delta S_{min}\,t_{sample}/\tau_{grp} = 25~{\rm Jy}$. Selected
GRPs are shown in the Fig.~\ref{grps}.

About $2.4 \times 10^6$ periods were observed with EPOS. The results
presented here are based on a one-third of all observed rotations only.
This restriction was
due to insufficient S/N in the remaining data. Because of the Crab
pulsar's weak signal at  8.35~GHz and the strong nebula background,
ordinary single pulses were not observable with Effelsberg at that frequency. 
It takes about 20 min integration to assemble a mean profile at that frequency.    
The giant pulses come in short episodes, about 5 to 20 minutes in duration,
and appear extremely prominent during such burst phases.

In our subsequent analysis, we set a threshold level of 5 rms  =  125 Jy on
the sum of RHC and LHC for the same phase bin to count as a detection of 
a giant pulse. More than 1300 giant pulses were detected that way
(Fig.~\ref{grps}) and their arrival phases were computed by using
the TEMPO{\footnote{http://pulsar.princeton.edu/tempo}}
pulsar timing package. The data were aligned using the current Jodrell Bank
timing model. They were  found to be a perfect match to the time of 
arrivals (TOAs) obtained at Jodrell Bank before and after the Effelsberg
observing session. We found that the giant pulses
occur at all those phases where the radio components of the Crab
pulsar emission are observed.

\section{Polarization}
The average polarization characteristics of ~900 GRPs in some aspects
are similar to already published high radio frequencies observations,
but in some aspects we do observe significant differences.
For all authors (this work, \cite{Moffett1999} and \cite{Kar2004})
relative offset of the position angle (P.A.) between IP and HFCs
is on the level of
40-45 degree. However, the P.A. for the IP, and at the same time
for the HFCs differ for all authors. They are as follow,
for the IP: 30$^\circ$, 0$^\circ$, -30$^\circ$, and for the HFCs:
60$^\circ$ to 70$^\circ$, 45$^\circ$, 0$^\circ$ according
to \cite{Moffett1999}, \cite{Kar2004} and this work, respectively.
Moreover, there is some discrepancy in the degree of polarization.
Almost 100\% linear polarization of all three components has been
derived by \citet{Kar2004}, whereas in \cite{Moffett1999} the IP
is polarized in only 50\% and HFCs in 80-90\%. In our
work all components are polarized at the same level of 70-80\%.
This may be due to the time varying contribution of the nebula
\cite{Rankin1988} to the rotation measure of the pulsar
(\cite{Rankin1988}: RM=-43 rad m$^{-2}$,
\cite{Moffett1999}: RM=-46.9 rad m$^{-2}$, 
\cite{Weisberg2004}: RM=-58 rad m$^{-2}$).
No abrupt sweeps in P.A. are found within pulse components.

The study of radio polarization has not
improved essentially our knowledge of the emission and
magnetic field geometry: e.g., by fitting the rotation vector model
to the radio polarization data uttery opposite conclusions had been
reached (\cite{Moffett1999}, \cite{Kar2004}).
\citet{Moffett1999} found that the angle between the rotation and
magnetic axes is $\alpha=56^{\circ}$, while \citet{Kar2004}
got nearly aligned rotator with $\alpha=4^{\circ}\pm1^{\circ}$.
They obtained different results for the impact angle as well.

\begin{figure}
  \includegraphics[height=.54\textheight]{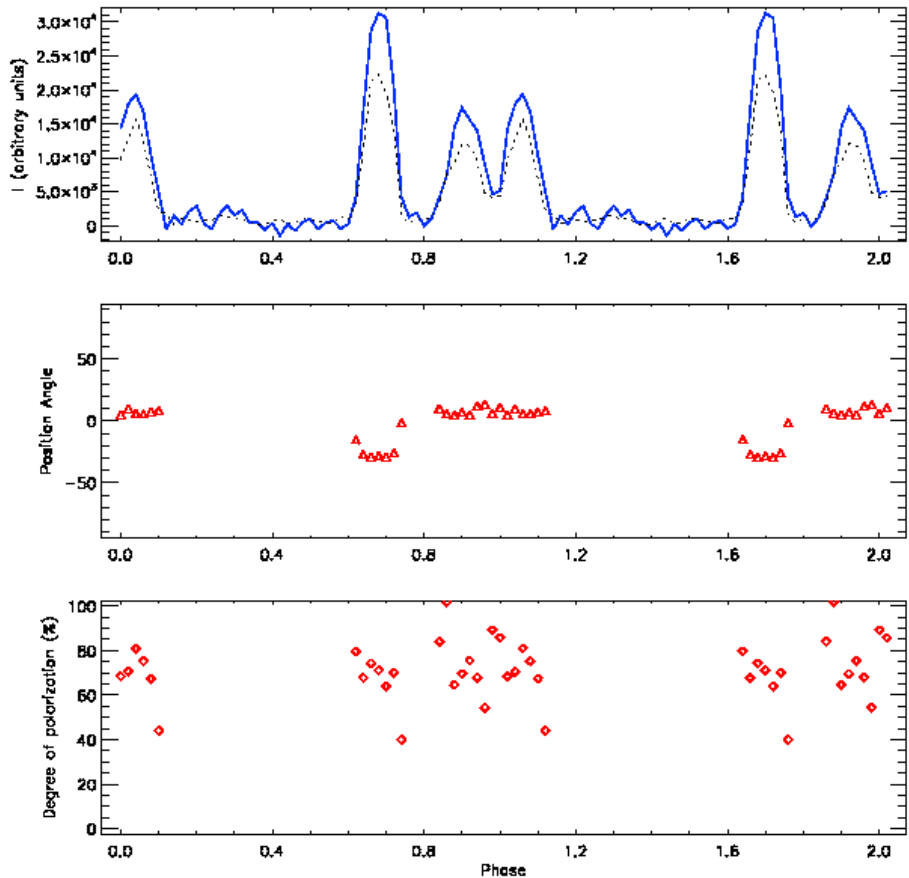}
  \caption{The average intensity profile (\emph{top}),
the position angle (after Faraday correction, \emph{middle}),
and the degree of polarization (\emph{bottom}) of the 874 Crab's
GRPs recorded at 8.35~GHz. Location of the Crab's pulsar 
radio components:
HFC2: 0.05, precursor: 0.2,
MP: 0.3, IP: 0.68, HFC1: 0.9.}
  \label{av_pol}
\end{figure}

\begin{figure}
  \includegraphics[height=.55\textheight]{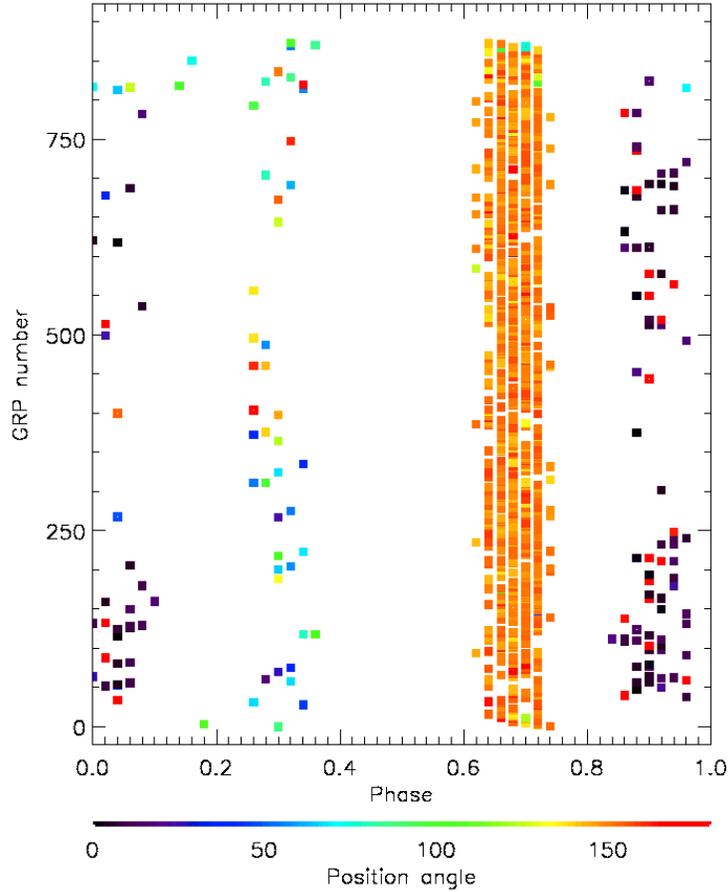}
  \caption{The position angle of selected 874 GRPs as a function of rotational
phase of the Crab pulsar. P.A. is plotted only for points above
$4.5\sigma$ of the off-pulse noise. Location of the Crab's pulsar radio
components: HFC2: 0.05, precursor: 0.2, MP: 0.3, IP: 0.68, HFC1: 0.9.
(Full colour version can be found at the astro-ph service).}
  \label{sing_pol}
\end{figure}
In the Fig.~\ref{sing_pol} the position angle of selected 874~GRPs as
a function of rotational phase of the Crab pulsar is shown. In
these polarization characteristics of single pulses we do not
observe any sudden changes of the P.A. within the separated components,
as it has been noticed for PSR B0329+54 by \citet{Edwards2004}.

\section{GRPs distribution}
For all giant pulses, i.e. regardless of their phases, the histogram
of their peak strengths at 8.35 GHz can be described by a power-law with a
slope $\sim -3.34 \pm 0.19$ (\cite{Jessner2005}) . This result is consistent
with the results obtained by others (e.g. \cite{Lundgren1995}).
Proportional contribution of a number of GRPs to four different 
phase components is as follow: IP - 80\%, MP - 9\%, HFC1 - 7\%,
and HFC2 - 4\%. Because of the limited statistics we could make a model
$(S/N)^{\alpha}$ fit only for the IP component, where a power-law index
$\sim -3.13 \pm 0.22$ was found, and it is consistent with
the value of $\sim -2.9$ presented by \citet{Cordes2004}.
The MPs from the Crab pulsar at 146~MHz are distributed according
to a power-law with the exponent of -2.5, whereas the IPs
with -2.8 \cite{Argyle1972}. For comparison at 800 MHz the distribution
(regardless of the phases of GRPs) has a slope of -3.46 \cite{Lundgren1995}.
It should be noticed that at low radio frequency the main
contribution to the number of GRPs comes from main pulses.
So far, there is no evidence that the distributions of numbers
of GRPs for the HFC1 and HFC2 differ from each other. However, their slopes
seems to be steeper than the slope of the same distribution for the
IP component.

\section{Conclusions}
The results obtained so far by us suggest that the physical conditions 
in the regions responsible for HFCs emission might be  similar to those 
in the main pulse (MP) and interpulse (IP) emission regions.
This idea is supported not only by our detection of GRPs phenomenon 
at HFCs phases, but also by their polarization characteristics.
Still, the origin of high frequency components is an open question.
One of the proposition can be an inward emission \cite{Cheng2000} 
that for the Crab light curve produce two, properly placed and 
of good separation, additional peaks.

So far, it has been thought that the flux density distribution of giant pulses
has power-law statistics and that they appear associated with non-thermal high
energy emission (PSR B0531+21, PSR B1937+21, PSR B1821-24, and PSR B0540-69).
The characteristics of some pulsars, e.g. PSR B1133+16 \cite{Kramer2003}, and the discovery of the Crab's giant radio pulses at phases where no high energy emission is known, might rule out this definition. Other observations however seem to support the relation of high energy and radio emissions. Recently it was found that there is a correlation between X-ray and radio pulses for Vela \cite{Donovan2004}, whereas Shearer et al.
\cite{Shearer2003} have detected a correlation between optical
emission and GRPs emission. They found that optical pulses coincident
with radio giant pulses were of about 3\% brighter on average.


\begin{theacknowledgments}
Aga S\l{}owikowska was supported by grant KBN 2P03D.004.24. She 
would like to thank Bronek Rudak for his very bright suggestions.
\end{theacknowledgments}





\begin{thebibliography}{17}
\expandafter\ifx\csname natexlab\endcsname\relax\def\natexlab#1{#1}\fi
\providecommand{\enquote}[1]{``#1''}
\expandafter\ifx\csname url\endcsname\relax
  \def\url#1{\texttt{#1}}\fi
\expandafter\ifx\csname urlprefix\endcsname\relax\def\urlprefix{URL }\fi
\providecommand{\eprint}[2][]{\url{#2}}

\bibitem[{Staelin} and {Reifenstein}(1968)]{Staelin1968}
D.~H. {Staelin}, and E.~C. {Reifenstein}, \emph{Science} \textbf{162},
  1481--1483 (1968).

\bibitem[{Gower} and {Argyle}(1972)]{Gower1972}
J.~F.~R. {Gower}, and E.~{Argyle}, \emph{ApJ} \textbf{171}, L23--L26 (1972).

\bibitem[{Lundgren} et~al.(1995)]{Lundgren1995}
S.~C. {Lundgren}, J.~M. {Cordes}, M.~{Ulmer}, S.~M. {Matz}, S.~{Lomatch}, R.~S.
  {Foster}, and T.~{Hankins}, \emph{ApJ} \textbf{453}, 433--445 (1995).

\bibitem[{Sallmen} et~al.(1999)]{Sallmen1999}
S.~{Sallmen}, D.~C. {Backer}, T.~H. {Hankins}, D.~{Moffett}, and S.~{Lundgren},
  \emph{ApJ} \textbf{517}, 460--471 (1999).

\bibitem[{Cordes} et~al.(2004)]{Cordes2004}
J.~M. {Cordes}, N.~D.~R. {Bhat}, T.~H. {Hankins}, M.~A. {McLaughlin}, and
  J.~{Kern}, \emph{ApJ} \textbf{612}, 375--388 (2004).

\bibitem[{Jessner} et~al.(2005)]{Jessner2005}
A.~{Jessner}, A.~{S{\l}owikowska}, B.~{Klein}, H.~{Lesch}, C.~H. {Jaroschek},
  G.~{Kanbach}, and T.~H. {Hankins}, \emph{Advances in Space Research}
  \textbf{35}, 1166--1171 (2005).

\bibitem[{Moffett} and {Hankins}(1996)]{Moffett1996}
D.~A. {Moffett}, and T.~H. {Hankins}, \emph{ApJ} \textbf{468}, 779--783
  (1996).

\bibitem[{Moffett} and {Hankins}(1999)]{Moffett1999}
D.~A. {Moffett}, and T.~H. {Hankins}, \emph{ApJ} \textbf{522}, 1046--1052
  (1999).

\bibitem[{Karastergiou} et~al.(2004)]{Kar2004}
A.~{Karastergiou}, A.~{Jessner}, and R.~{Wielebinski}, \enquote{{High-frequency
  Polarimetric Observations of the Crab Pulsar},} in \emph{IAU Symposium},
  2004, pp. 329--330.

\bibitem[{Rankin} et~al.(1988)]{Rankin1988}
J.~M. {Rankin}, D.~B. {Campbell}, R.~B. {Isaacman}, and R.~R. {Payne},
  \emph{A\&A} \textbf{202}, 166--172 (1988).

\bibitem[{Weisberg} et~al.(2004)]{Weisberg2004}
J.~M. {Weisberg}, J.~M. {Cordes}, B.~{Kuan}, K.~E. {Devine}, J.~T. {Green}, and
  D.~C. {Backer}, \emph{ApJS} \textbf{150}, 317--341 (2004).

\bibitem[{Edwards} and {Stappers}(2004)]{Edwards2004}
R.~T. {Edwards}, and B.~W. {Stappers}, \emph{A\&A} \textbf{421}, 681--691
  (2004).

\bibitem[{Argyle} and {Gower}(1972)]{Argyle1972}
E.~{Argyle}, and J.~F.~R. {Gower}, \emph{ApJ} \textbf{175}, L89--L91 (1972).

\bibitem[{Cheng} et~al.(2000)]{Cheng2000}
K.~S. {Cheng}, M.~{Ruderman}, and L.~{Zhang}, \emph{ApJ} \textbf{537},
  964--976 (2000).

\bibitem[{Kramer} et~al.(2003)]{Kramer2003}
M.~{Kramer}, A.~{Karastergiou}, Y.~{Gupta}, S.~{Johnston}, N.~D.~R. {Bhat}, and
  A.~G. {Lyne}, \emph{A\&A} \textbf{407}, 655--668 (2003).

\bibitem[{Donovan} et~al.(2004)]{Donovan2004}
J.~{Donovan}, A.~{Lommen}, Z.~{Arzoumanian}, A.~{Harding}, M.~{Strickman},
  C.~{Gwinn}, R.~{Dodson}, P.~{McCulloch}, and D.~{Moffett},
  \enquote{{Correlations Between X-ray and Radio Pulses in Vela},} in \emph{IAU
  Symposium}, 2004, pp. 335--336.

\bibitem[{Shearer} et~al.(2003)]{Shearer2003}
A.~{Shearer}, B.~{Stappers}, P.~{O'Connor}, A.~{Golden}, R.~{Strom},
  M.~{Redfern}, and O.~{Ryan}, \emph{Science} \textbf{301}, 493--495 (2003).

\end{thebibliography}

\IfFileExists{\jobname.bbl}{}
 {\typeout{}
  \typeout{******************************************}
  \typeout{** Please run "bibtex \jobname" to optain}
  \typeout{** the bibliography and then re-run LaTeX}
  \typeout{** twice to fix the references!}
  \typeout{******************************************}
  \typeout{}
 }

\end{document}